# Citation distribution of individual scientist: approximations of stretch exponential distribution with power law tails

O.S. Garanina and M.Yu. Romanovsky[1]

[1]*slon@kapella.gpi.ru*
A.M.Prokhorov General Physics Institute of RAS, Vavilov str., 38, 119991 Moscow (Russia)

**Abstract**
A multi-parametric family of stretch exponential distributions with various power law tails is introduced and is shown to describe adequately the empirical distributions of scientific citation of individual authors. The four-parametric families are characterized by a normalization coefficient in the exponential part, the power exponent in the power-law asymptotic part, and the coefficient for the transition between the above two parts. The distribution of papers of individual scientist over citations of these papers is studied. Scientists are selected via total number of citations in three ranges: $10^2$-$10^3$, $10^3$-$10^4$, and $10^4$-$10^5$ of total citations. We study these intervals for physicists in ISI Web of Knowledge. The scientists who started their scientific publications after 1980 were taken into consideration only. It is detected that the power coefficient in the stretch exponent starts from one for low-cited authors and has to trend to smaller values for scientists with large number of citation. At the same time, the power coefficient in tail drops for large-cited authors.
One possible explanation for the origin of the stretch-exponential distribution for citation of individual author is done.

**Conference topic**
Citation and co-citation analysis

**Introduction**
The discussion how citations of individual author are distributed has a long history climbed even to E.Garfild (Garfield 1955). In general, there are two points of views on this present: the distribution of papers of each scientist is so-called stretch exponent $W \sim \exp(-x^\alpha/T)$ where $x$ is a number of citations, $T$ is some normalization, $\alpha$ is the power exponent coefficient (Redner 1998, Laherrere&Sornette 1998). Usually $\alpha$ considered as 0,3-0,5 (Redner 1998, Iglesias&Pecharroman 2006). Slightly more complicated distribution was introduced by Tsallis&de Albuquerque 2000) (see also Wallace et.al. 2009, Anastasiadis et.al. 2010). Note that the first use of stretch exponential distributions was done by M.Subbotin (1923) for astrophysical purposes.
The second point is that the above distribution has power-law (Pareto, Zipf) character, i.e. W ~ $x^{\beta}$ where $\beta$ is the power (Silagadze 1999, Vazquez 2001, Lehmann et.al. 2003, Perc 2010, Rodríguez-Navarro 2011). Often, this dependence is treated as the asymptote (tail) of distribution for comparably large $x$. In this case, the main body is considered as log-normal (Redner 2005, Stanley 2010, Bommarito&Katz 2010, Chatterjee et.al. 2014). It should be noted that there are more complicated models of citation distribution.
The idea of our work is to consider the citation distribution of individual scientists taking into account that the distributions for "various-ranking" scientists can be different. As well it is interesting to join the above stretch-exponential distributions and power-law ones: observation of tails of citation distribution of individual science often demonstrates a presence of small number of extremely-high cited articles while other articles of considered scientist can be cited much more moderate. From this point of view, the consideration of big massive of data of citation of large set of authors (like in (Redner 1998) etc.) provides rough enough re-sults. Thus, we concentrate on analysis of citation distribution of individual scientists, taking into account some difference in total number of citations of them. The cumulative distribution of number of articles with some or larger number of citations will be analyzed.

Of course, the proposed approach is rough enough since does not take into account the co-authoring of cited articles. Authors think that it should be considered in next studies (may be, not only by us) in case of wide scientific interest.

The descriptive model based on our previous works for tailed distributions: Gauss for stock return distributions (Romanovsky&Vidov 2011), and exponential Boltzmann distribution for new car sells, incomes and weights (Romanovsky&Garanina 2015). Authors do not know consistently introduced mathematical formulae for distributions with exponential main part and power law asymptote.

**Multi-parametric family of curves with stretch exponential main part and power law tail**

To define the general form of the desired distribution, one may proceed from the results presented in (Romanovsky&Vidov 2011) as a starting point. According to (Romanovsky&Vidov 2011), that the sum of a large quantity $N$ of random values similarly distributed with the probability density function (PDF) of the Student's (generally, non-integer) type $\sim z_0^{2\beta}/(z_0^2 + f^2)^{2\beta}$ has the distribution of the Gaussian form for comparably small values of fluctuations $f$:

$$W_G(f) \approx \frac{1}{\sqrt{\pi}} \exp(-f^2)$$

and $\sim 1/f^{2\beta}$ for large $f$ ($z_0$ being a normalization constant, the sum is treated as random walks in (Romanovsky&Vidov 2011)). The obvious mathematical generalization to get the exponential part with power-law tail is to perform the transformation $f^2 \rightarrow R/T$ (here $T$ can be interpreted as an effective "temperature"). Upon switching from parameters $N$, $z_0$, $\beta$ to parameters $\theta$, $T$, $\sigma\beta$, the transformation yields the curve with the stretch exponential main part and a transition to power law at the tail in an explicit form of a PDF (Romanovsky&Garanina 2015):

$$W_{T(\sigma\beta)\theta}(R) = \frac{1}{\sqrt{\pi T}} \int_0^\infty \cos(xR^\sigma) \left\{ \frac{2}{\Gamma(\beta-1/2)} \left[ (\beta - 3/2) \frac{xT}{4\theta} \right]^{\beta/2 - 1/4} K_{\beta-1/2} \left[ \sqrt{(\beta - 3/2) \frac{xT}{4\theta}} \right] \right\}^\theta dx \quad (1)$$

Here $R$ is variable, $\Gamma$ is the gamma-function, $K_{\beta-1/2}$ is the modified Bessel function of the 2nd kind (also known as ''McDonald function'').

The approximation of Eq. (1) for comparably small $R$ (up to several units of $T^{1/2\sigma}$) is easily reduced to a dependence on parameter $T$ only

$$W_T(R) \cong \frac{1}{T} \exp\left(-\frac{R^{2\sigma}}{T}\right) \quad (2)$$

The general drop off law for $W_{T\beta\theta}$ in the case of large $R$ is $R^{-\beta\sigma}$. The parameter $\theta$ describes transition among (stretch) exponential and power-law part of (1). This transition goes under larger $R$ (and smaller values of $W_{T(\sigma\beta)\theta}$) under larger values of $\theta$.

To obtain a general form of $W$, note that

$$I_\beta(x) = \frac{2}{\Gamma(\beta-1/2)} \left[ (\beta - 3/2) \frac{xT}{4\theta} \right]^{\beta/2 - 1/4} K_{\beta-1/2}\left[ x\sqrt{(\beta - 3/2) \frac{T}{\theta}} \right], \quad (3)$$

It is easy to see that it is a monotonic function of $\beta$. Indeed, if $\nu = \mu + 1$, one finds, considering the rule for modified Bessel functions of the 2nd kind, that the ratio $I_\mu(x)/I_\nu(x)$ becomes

$$\frac{I_\mu(y)}{I_\nu(y)} = \frac{K_{\mu+1/2}(y) - K_{\mu-3/2}(y)}{K_{\mu+1/2}(y)} = 1 - \frac{K_{\mu-3/2}(y)}{K_{\mu+1/2}(y)} < 1$$

Furthermore, $\forall \eta : \nu > \eta > \mu$, and one finds that $I_\nu > I_\eta > I_\mu$. Thus, it is not necessary to investigate (1,3) with an arbitrary $\beta$. It is enough to consider the integer $\beta = 2, 3, \ldots$, while integrals with intermediate $\beta$ will be ''locked'' among integrals with neighboring integers $\beta$ that are expressed by means of elementary functions. Then $n=\beta-1$,

$$K_{\beta-1/2}\left[x\sqrt{(\beta-3/2)\frac{T}{\theta}}\right] = K_{n+1/2} = \sqrt{\frac{\pi}{2x\sqrt{(\beta-3/2)\frac{T}{\theta}}}}\sum_{k=0}^{n}\frac{(n+k)!}{k!(n-k)!\left[2x\sqrt{(\beta-3/2)\frac{T}{\theta}}\right]^k} \quad (4)$$

The three functions $W_{T(\sigma\beta)\theta}$ for $\sigma\beta=2, 1, 0.8$ are:

$$W_{T(\sigma\beta)\theta}(R) = \frac{1}{\sqrt{\pi T}}\int_0^\infty \cos(xR^{\sigma\beta})\exp\left(-x\sqrt{\frac{\theta T}{2}}\right)\left(1+x\sqrt{\frac{T}{2\theta}}\right)^\theta dx \quad (5)$$

We used here the simplest form of the function (1) for $\beta=2$ for the following approximations of empirical data. The functions $W_{T(\sigma\beta)\theta}$ for $\sigma= 0.5, 0.25, 0.2$ are shown in Fig.1. It is seen well-coincidence of general functions with corresponding approximation exponents for comparably small values of variable $R$.

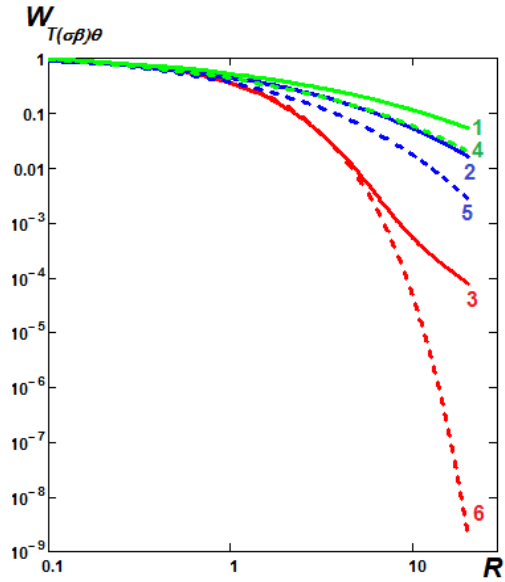

**Fig.1. Functions $W_{T(\sigma\beta)\theta}$ for $\beta=2$ and $\sigma=0.5$ (curve 3), $\sigma=0.25$ (curve 2), $\sigma=0.2$ (curve 1) for comparably small $R$. The straight lines (4-6) are exponents $exp(-R^{2\sigma}/T)$ for $\sigma=1, 0.5, 0.4$ respectively. Here $T=1$, $\theta=300$.**

For large $R$, these functions drop off as $R^{-2}$, $R^{-1}$, $R^{-0.8}$ respectively (see Fig.2):

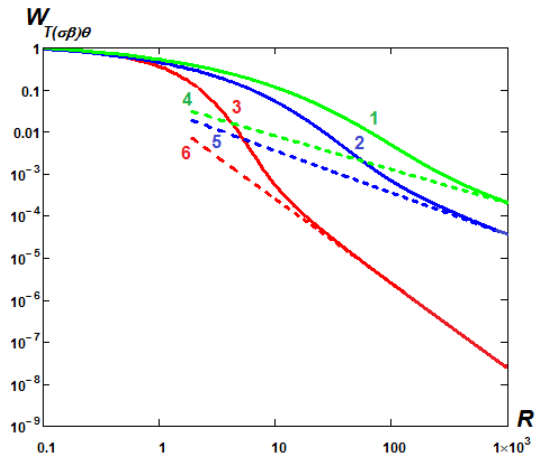

**Fig.2.** Functions $W_{T(\sigma\beta)\theta}$ for the same $\beta$ and $\sigma$ (curves 3-1) as on Fig.1. Hyperboles $R^{-\beta\sigma}$ (straight lines 6-4 on double-logarithmic plot) have $\sigma$=0.5, 0.25, and 0.2 (curve 4) respectively. Parameters $T$, $\theta$ are the same as on Fig.1.

Thus the introduced function (1) well-describes the stretch exponent for small (and moderate) values of argument, and provides power-law asymptotes for large $R$. We used these functions in the next section.

**Distribution of citation of individual authors**

One was found that distributions of citation of individual authors are different. It can be expected due to, for example "Matthew effect" (see Bonitz et.al. 1997, Bonitz&Scharnhorst 2001, Stanley 2010). One may expect that scientists with total number of citation in range $10^2$-$10^3$, $10^3$-$10^4$, and $10^4$-$10^5$ have different distributions of citation. Let us call the scientists with total number of citations in these ranges as the "first-type scientist", etc. We study these intervals for physicists in ISI Web of Knowledge. The scientists who started their scientific publications after 1980 were taken into consideration only. One was taken 20 scientists for the first two ranges, and several scientists for the third. Typical examples of citation distributions are presented below on Figs. 3-5.

On Fig. 3, the cumulative citation distribution (i.e. the number of articles with citations larger than the value $R$) for experienced scientists with total number of citations in the first range $10^2$-$10^3$ is presented:

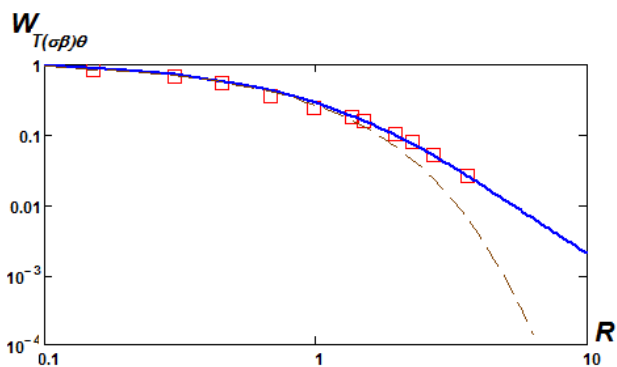

**Fig. 3.** The distribution of articles over citations for the first-type scientist. Open squares are empirical points, the solid curve is $W_{T(\sigma\beta)\theta}$ (5) for $\beta$=2, $\sigma$=0.5, $T$=6.5, $\theta$=10, dashed line is an exponent (2) with $\sigma$=0.5, $T$=6.5.

The function $W_{T(\sigma\beta)\theta}$ on Fig.3 is normalized on total number of articles of the first-type scientists in ISI Web of Knowledge. The variable $R$ is the number of citations normalized on $T$ that is the mean citation of this author. It is seen that the function $W_{T(\sigma\beta)\theta}$ (5) well describes the empirical data, the clear difference from the exponent (2) is on-site. At the same time, the total exit on the asymptotic curve ~ $R^{-2}$ does not realize. The last was observed for other-types scientists.

The citation distribution of the second-type scientist (this is a range of world well-known person) is demonstrated on Fig. 4:

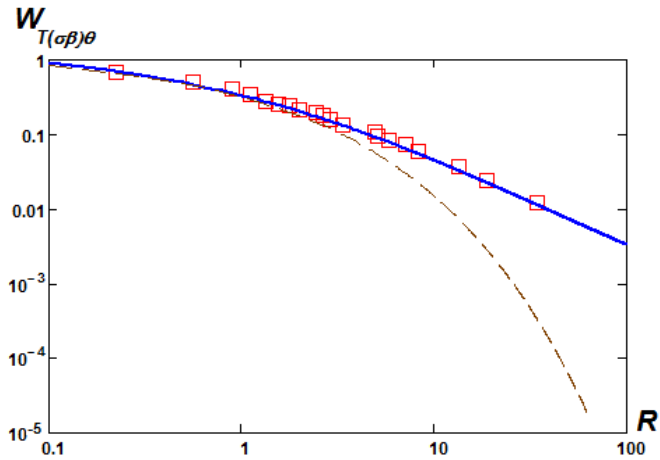

**Fig.4. The distribution of articles over citations for the second-type scientist. Open squares are empirical points, the solid curve is $W_{T(\sigma\beta)\theta}$ (5) for $\beta=2$, $\sigma=0.25$, $T=47.4$, $\theta=5$, dashed line is an exponent (2) with $\sigma=0.25$, $T=46$.**

The normalization of $W_{T(\sigma\beta)\theta}$ on Fig.4 was on total number of articles also. Indeed, the variable $R$ is normalized now on $T^{2\sigma} = (47.4)^{2\sigma} = 6.9$. The "difference" between empirical data as well as function (5) with pure stretch exponent $exp(-R^{1/2}/T)$ is larger than on Fig.3 for the first-type scientist. The total exit on the asymptotic curve ~ $R^{-1}$ does not realize also.

The citation distribution of the third-type scientist (this is a range of Nobel Prize winner) is demonstrated on Fig. 5:

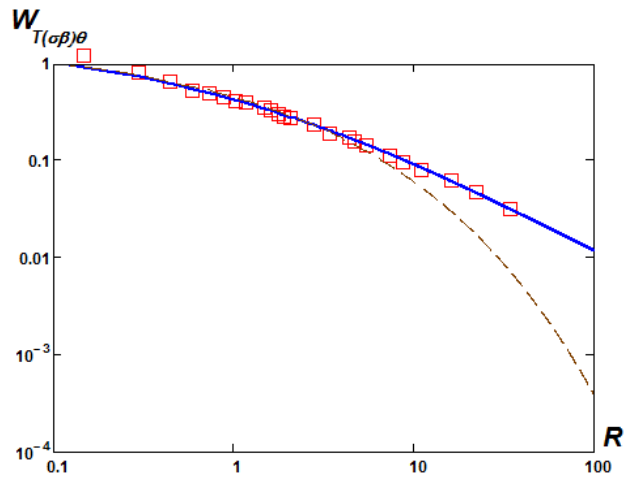

**Fig.5. The distribution of articles over citations for the thhird-type scientist. Open squares are empirical points, the solid curve is $W_{T(\sigma\beta)\theta}$ (5) for $\beta=2$, $\sigma=0.2$, $T=340$, $\theta=5$, dashed line is an exponent (2) with $\sigma=0.2$, $T=340$.**

The normalization of $W_{T(\sigma\beta)\theta}$ on Fig.5 is the same, the variable $R$ is normalized now on $T^{2\sigma} = 340^{2\sigma} = 10.3$. It is interesting that all values $T^{2\sigma}$ for all three-types scientists close to each other and may characterize the citation distribution of individual scientist.

**Explanation attempt**

Let us try to explain the appearance of stretch exponents in cumulative distribution of such random values like citations. We start from the standard exponential distribution

$$W_1 = \exp(-x) \qquad (6)$$

Where we used normalization $T=1$ to simplify the following expressions. Note that distributions like (6) arise often in problems of two-body exchange in process of body collisions (in this case, there is two-body exchange of energy, see Landau&Lifshitz 1980), in money distribution (see Dragulescu&Yakovenko 2000).

How these calculations can be "translate" into the language of citations? The first cause of citation of some article is the scientific results of this article. Since the author who can potentially cite the above article may find or not find this article, the process of citation due to the scientific significance looks like the two-body exchange (of information in this case) and is provided by distribution (6). Thus it may impress that the basic citation cumulative distribution arise due to the scientific significance of the article and looks like (6).

There are clear additional independent causes for citation. One of them is the name of author (or one of authors in case of co-authoring) of potentially cited article. It may be the name of scientist in the group that works in the same area of science studied the author of cited paper, there arise another causes to cite some scientist. Since this scientist may also be chosen randomly in the process of information exchange, the probability distribution to cite this scientist looks like (6) also. If now the citation is realized due to two causes: by scientific significance and cited article author, the random value of such citation is the factor of two random values characterized by distribution (6).

Since the causes for citation are independent, they can be considered as some coordinates. For two cases, they are above "scientific significance" and "author's name". The variation of these coordinates here are from small to large scientific significance and from large to small reputation of cited scientist. Thus the citation now is the two-dimension "square" procedure in comparison with "linear" for one cause (see Fig.6):

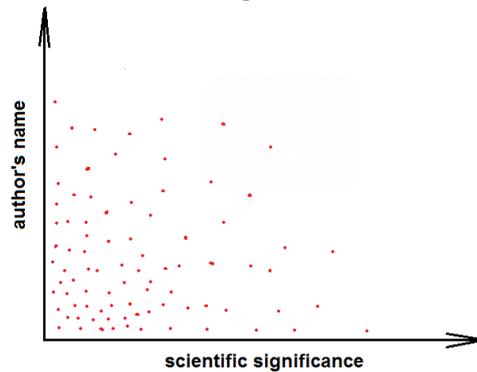

**Fig.6. The distribution of citations (dots) in two "coordinates": "scientific significance" and "author's name", see explanations in text**.

At the same time, we observe citation as principally one-dimension value: the citation either exists or does not exist. Therefore, all distributions (6) reduce to one dimension. The transformation of coordinates in (7) $x^2 \to y$ provides than for cumulative distribution function

$$W_2(y) = \exp(-\sqrt{y}) \tag{7}$$

i.e. the main part of stretch exponent (2) with $\sigma=0.25$. These stretch-exponents distributions were observed by us and described in the chapter of this paper "Distribution of citation of individual authors".

The same procedure in case of three clearly existed "coordinates" provides cumulative distribution

$$W_3(y) = \exp(-\sqrt[3]{y}) \tag{8}$$

etc.

Of course, all above speculations may be apply to pure stretch exponential distributions directly. However the same conduction for power-law tailed stretch exponential distributions should take into consideration the power exponents in tails for original distributions of "scientific significance" etc., and needs the volumetric calculations.

## Conclusion

The 4-parametric family of functions representing the stretch exponential distribution for small and medium values of the argument combined with a power-law asymptotic tail, along with various transitions between these two parts, is introduced. These functions are demonstrated good fitting of the available empirical data for the cumulative distribution of articles of individual scientist over their citations.

Abstracting from the co-authoring of cited paper, one may conclude that these cumulative distributions of papers of individual author versus their citations have character of stretch exponent for small and moderate values of citations, and power-law form for asymptotic part. It looks that the "power of stretch", i.e. the introduced coefficient $\sigma$ depends on the total number of citations, moreover, this coefficient starts from ½ (i.e. distributions start from normal exponent) and becomes smaller with an increase of the total number of citations. The power-law force becomes smaller in return.

The first attempt to explain the "main body" of distributions (stretch exponents) is provided.

## Acknowledgements

The paper is support by RFBR grant 13-07-00672.

## References

Anastasiadis, A.D., de Albuquerque, M.P., Mussi, D.B. (2010). Tsallis q-exponential describes the distribution of scientific citations - a new characterization of the impact. *Scientometrics,* 83, 205–218.

Bommarito, M.J., Katz, D.M. (2010). A mathematical approach to the study of the united states code. *Physica A*, 389. 4195–4200.

Bonitz, M., Brukner, E., Scharnhorst, A. (1997). Characteristics and impact of Matthew effect for countries. *Scientometrics*, 40, 407-422.

Bonitz, M., Scharnhorst, A. (2001). Competition in science and the Matthew core journals. *Scientometrics*, 51, 37-54.

Chatterjee, A., Asim Ghosh, A., Chakrabarti, B.K. (2014). Universality of citation distributions for academic institutions and journals. http://arxiv.org/abs/1409.8029v1

Garfield, E. (1955). Citation indexes for science: A new dimension in documentation through association of ideas. *Science*, 122. 108–111.

Dragulescu, A., and Yakovenko, V.M. (2000). Statistical mechanics of money. *The European Physical Journal B*, 17, 723–729.

Iglesias, J.E., Pecharroman, C. (2006). Scaling the h-Index for Different Scientific ISI Fields. Online: http://arxiv.org/ftp/physics/papers/0607/0607224.pdf

Laherrere, J., Sornette, D. (1998). Stretched exponential distributions in nature and economy: "fat tails" with characteristic scales. *The European Physical Journal B*, 2, 525-539.

Landau, L.D., Lifshitz, E.M. (1980). Statistical Physics. Amsterdam: Butterworth-Heinemann; 3rd edition.

Lehmann, S., Lautrup, B., Jackson, A.D. (2003). Citation networks in high energy physics. *Physics Review E,* 68. 026113.

Perc, M. (2010). Zipf's law and log-normal distributions in measures of scientific output across fields and institutions: 40 years of slovenia's research as an example. *J Informetrics,* 4, 358–364.

Petersen, A.M., Fengzhong Wang, and Stanley, H.E. (2010). Methods for measuring the citations and productivity of scientists across time and discipline. *Physics Review E,* 81, 036114.

Redner, S. (1998). How popular is your paper? An empirical study of the citation distribution. *The European Physical Journal B,* 4, 131-134.

Redner, S. (2005). Citation Statistics from 110 Years of Physical Review. *Physics Today,* 58. 49–54.

Rodríguez-Navarro, A. (2011). A simple index for the high-citation tail of citation distribution to quantify research performance in countries and institutions. *PLoS ONE,* 6, e20510.

Romanovsky, M.Yu., Vidov, P.V. (2011). Analytical representation of stock and stock-indexes returns: Non-Gaussian random walks with various jump laws. *Physica A*, 390, 3794–3805.

Romanovsky, M.Yu., Garanina, O.S. (2015). New multi-parametric analytical approximations of exponential distribution with power law tails for new cars sells and other applications. *Physica A*, 427, 1-9.

Silagadze, Z.K. (1999). Citations and the Zipf-Mandelbrot's law. Online: http://arXiv.org/abs/physics/9901035v2

Subbotin, M.T. (1923). Law of Error Distribution. *Rec.Math. [Matem. Sbornik].,* 31, 296-300.

Tsallis, C., de Albuquerque, M.P. (2000). Are citations of scientific papers a case of nonextensivity? *The European Physical Journal B*, 13, 777-780.

Vazquez, A. (2001). Statistics of citation networks. *E-prints* arXiv:condmat/0105031.

Wallace, M.L., Larivière, V., Gingras, Y. (2009). Modeling a Century of Citation Distributions. *Journal of Informetrics*, 3, 296–303.